\documentclass[3p,12pt]{elsarticle}
\biboptions{sort&compress}
\usepackage{amssymb}
\usepackage{color,slashed,mathrsfs}
\usepackage{graphicx}

\definecolor{red}{rgb}{0.8,0,0}
\definecolor{violet}{rgb}{0.4,0,0.4}
\definecolor{green}{rgb}{0,0.5,0.0}
\definecolor{navy}{rgb}{0.0,0.0,0.6}
\definecolor{orange}{rgb}{0.8,0.2,0.0}
\definecolor{blue}{rgb}{0.3,0.0,0.8}

\begin{document}
\journal{}

\begin{frontmatter}
\title{Investigating the glue content of $f_1(1285)$}
\author{P. G. Moreira and M. L. L. da Silva}
\address{Instituto de F\'{\i}sica e Matem\'atica, Universidade Federal de
Pelotas, Caixa Postal 354, CEP 96010-090, Pelotas, RS, Brazil.}

\date{\today}

\begin{abstract}
The radiative decay of $f_1(1285)$ to vector mesons $\rho$ and $\phi$ is calculated in a
non-relativistic approach. The results obtained are compared to experimental data to
estimate the glue content of $f_1$. We find a small or nonexistent glue content when our
results are compared to the PDG data. However a considerable glue content for $f_1(1285)$ was
found when we compared our results to the CLAS data.
\end{abstract}

\begin{keyword}
 Radiative decay \sep glueball \sep hadron mixing 
\end{keyword}
\end{frontmatter}

\section{Introduction}

The light mesons sector has been studied over several decades and many properties of the 
mesons were found. However, there are still uncertainties about the properties of some of
these mesons. In particular, their structure are not yet fully established. For this reason,
this sector still arouses interest from both a theoretical and an experimental point of
view \cite{Battaglieri:2014gca}. The possible existence of exotic hadrons is also subject
of several studies \cite{Crede:2008vw,Mathieu:2008me,Ochs:2013gi}. Most of these exotic
mesons are not yet experimentally detected, but several theoretical and phenomenological
works predicts their existence and proprieties. The glueball is an exotic state predicted
theoretically but not yet experimentally observed. Glueballs and $q\bar{q}$ mesons states
can mixing strongly which brings numerous experimental difficulties to detect glueball
states doe to the interference of $q\bar{q}$ states.

The radiative decay of mesons can be used to establish their internal structure. For this 
reason, the meson in the final state must have a well established internal structure. This
idea was used to predict the radiative decay rates calculated in the context of a
non-relativistic model as a function of the mixing angles of glueball and meson states
\cite{Close:2002ky,Close:2002sf}. In their work Close et al have considered the $f_1(1285)$
as a member of the axial vector meson nonet \cite{Close:2002sf}.

The $f_1(1285)$, with quantum numbers $J^{PC} = 1^{++}$, mass $1281.9 \pm 0.5$ MeV and total
width $22.7 \pm 1.1$ MeV \cite{Tanabashi:2018oca}, is usually considered as a member of the
axial vector meson nonet. Recently, the photoproduction of $f_1(1285)$ was studied by
the CLAS Collaboration \cite{Dickson:2016gwc}, where the photoproduction differential cross
section and the branching ratio $\Gamma(\gamma \rho^0)/\Gamma(\eta\pi\pi) = 0.047 \pm 0.018$
were measured. Then they combine this data with PDG results and found $\Gamma[f_1(1285) \to
\gamma \rho^0] = 453 \pm 177$ keV which is in poor agreement with the PDG estimation of
$1203 \pm 280$ keV \cite{Tanabashi:2018oca}. The PDG Collaboration also shown the results
for the branching ratio of $\Gamma[f_1(1285) \to \gamma \phi] = 17,0 \pm 6.3$ keV.

From the point of view of its nature, the $f_1(1285)$ was mostly considered as a $q\bar{q}$
state, more precisely composed by $u$ and $d$ quarks. However, it was also predicted in the
literature \cite{Birkel:1995ct} a mixture of gluons in $f_1$ wave function. This mixture
was described in a form similar to the mixing in the scalar sector. On this way the 
resonances $f_1(1285)$, $f_1(1420)$ and $f_1(1510)$ were considered as a mixing of
$1/\sqrt{2} \mid u\bar{u} + d\bar{d} \rangle$, $\mid s\bar{s} \rangle$ and $\mid G \rangle$
states. This possibility was also mentioned by Kochelev et al in their paper about
$f_1(1285)$ photoproduction \cite{Kochelev:2009xz}.

In this work we calculate the radiative decay rates for $f_1(1285) \to \gamma \rho$ and
$f_1(1285) \to \gamma \phi$. The choice of $\rho$ and $\phi$ is doe to the $\rho$ is a
state of $u$ and $d$ quarks and $\phi$ is a state of $s$ quarks. Then we can use these
results and the experimental data available to estimate the quarks content of $f_1$ and
consequently the glue content of this resonance. In the next section we show the meson
confining potential and the relation of this potential with the parameters of the meson
wave function. In section III the non-relativistic radiative decay model is presented.
Our results are shown in section IV. The section V is about the Summary and Conclusions. 

\section{The confining potential}

In a non-relativistic constituent quark model, the meson is described as a quark-antiquark
system with masses $m_q$ and $m_{\bar{q}}$ respectively. The potential used here is the
Cornell potential which describe the behavior of the color interaction in the two asymptotic
limits \cite{KerenZur:2007vp}
\begin{eqnarray}
 V(r) = K r - \frac{4 \alpha_s}{3 r} + C \,.
 \label{pot}
\end{eqnarray}
where $K = 0.18$ GeV$^2$ and $\alpha_s = 0.39$ \cite{Close:2002ky,Ding:1998qr}. This
potential allow us to solve variationally the Schr\"odinger equation and connect the meson
masses and the parameters which are needed in the next section \cite{KerenZur:2007vp}
\begin{eqnarray}
 \left(m_q + m_{\bar{q}} + \frac{p^2}{2 \mu} + V(r)\right) \mid\Psi_M\rangle 
 = M \mid\Psi_M\rangle \, .
 \label{schr}
\end{eqnarray}
The wave functions are taken to be Gaussian multiplied by a polynomial. For $L=0$ the 
normalized wave function is given by
\begin{eqnarray}
 \Psi_M = \frac{2\beta_M^{3/2}}{\pi^{1/4}} \, e^{-\frac{\beta_M^2 r^2}{2}} \,
 {\cal Y}_{00}(\Omega)\,,
\end{eqnarray}
and for $L=1$ we have
\begin{eqnarray}
 \Psi_M = 2 \, \sqrt{\frac{3}{2}} \, \frac{\beta_M^{5/2}}{\pi^{1/4}} \, r \,
 e^{-\frac{\beta_M^2 r^2}{2}} \, {\cal Y}_{10}(\Omega)\,.
\end{eqnarray}
where $\beta_M$ is the parameter which will be obtained together with the mesons masses.

In the present work we consider two possibilities for the mesons states. The first
one is that only quarks $u$ and $d$ are part of the meson. On this way we consider
$m_q = m_{\bar{q}} = 0.33$ GeV. With these assumptions we obtain the mesons masses
and the $\beta_M$ parameters for $\rho(770)$ and $f_1(1285)$. The results obtained
are presented in Table \ref{tab1}.
\begin{table}[ht]
\caption{Parameters for mesons composed by $u$ and $d$ quarks. \label{tab1}}
\begin{center}
\begin{tabular}{ccc}
\hline
Meson & Mass (GeV) & $\beta_M$ (GeV) \\
\hline
$1^{--}$ & $0.758$ & $0.305$ \\
$1^{++}$ & $1.298$ & $0.270$ \\
\hline
\end{tabular}
\end{center}
\end{table}
A similar procedure is used for $s$ quarks composite mesons. In this case the quark
mass is $m_q = m_{\bar{q}} = 0.54$ GeV \cite{KerenZur:2007vp}. The results for
$\beta_M$ parameters and the masses are shown in Table \ref{tab2}.
\begin{table}[ht]
\caption{Parameters for mesons composed by $s$ quarks. \label{tab2}}
\begin{center}
\begin{tabular}{ccc}
\hline
Meson & Mass (GeV) & $\beta$ (GeV) \\
\hline
$1^{--}$ & $0.981$ & $0.371$ \\
$1^{++}$ & $1.463$ & $0.323$\\
\hline
\end{tabular}
\end{center}
\end{table}
In the next section the parameters found here will be used as an input for the
radiative decay model.

\section{Radiative Decay Model}

\begin{figure}[ht]
\begin{center}
\includegraphics[scale=0.3]{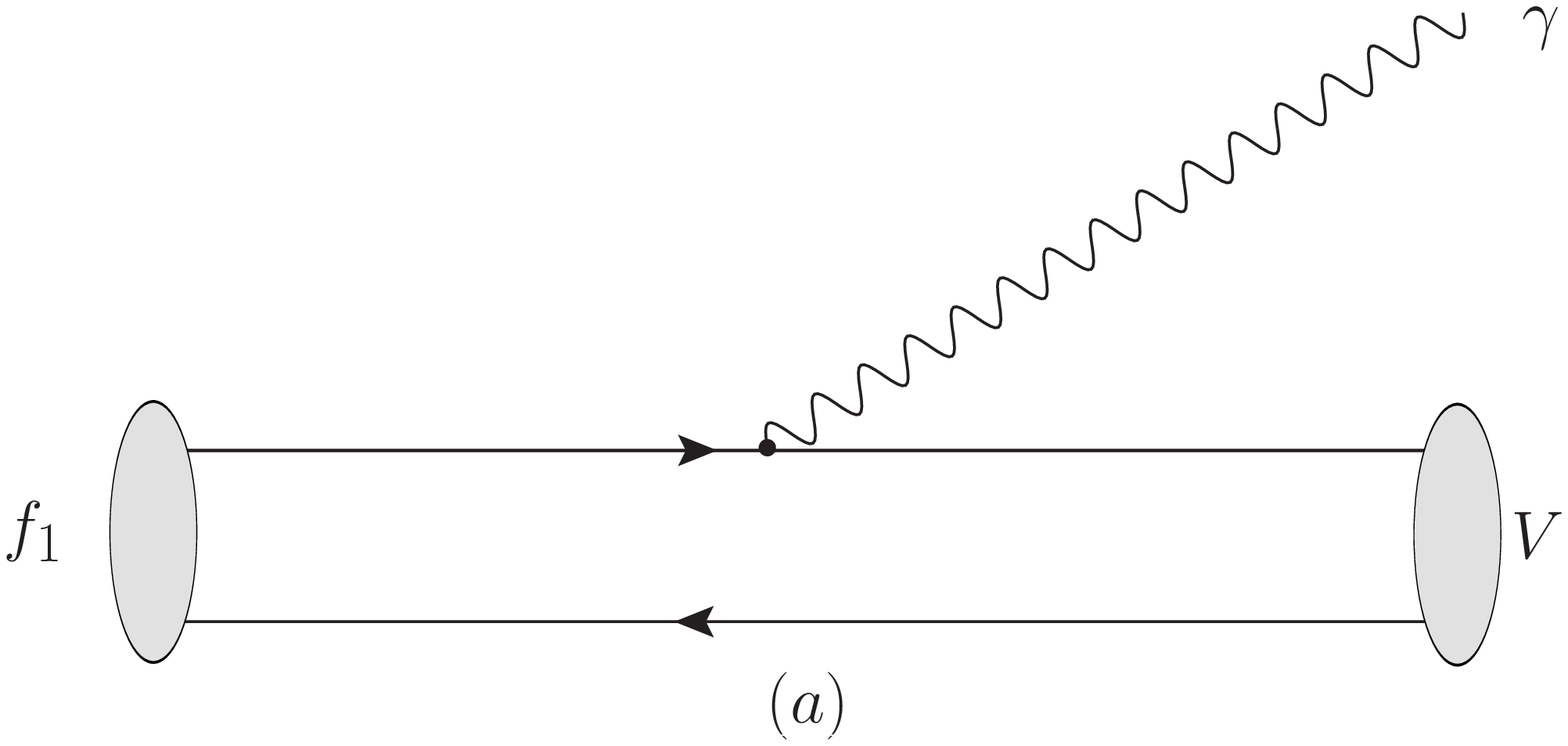} 
\vspace{0.8cm}
\includegraphics[scale=0.3]{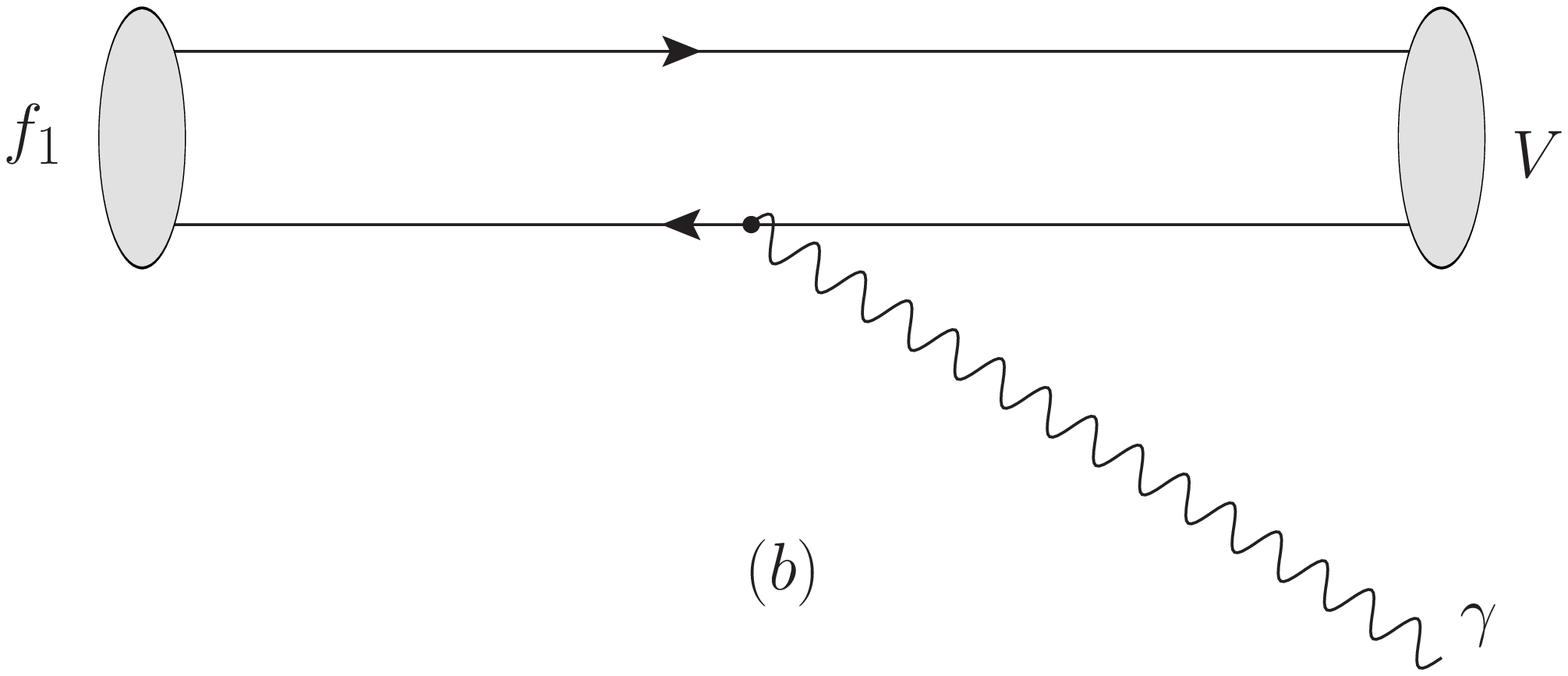}%
\caption{Diagrams representing the two possibilities for the radiative decay of
$f_1(1285)$.\label{diag1}}
\end{center}
\end{figure}

In this section we describe the formalism for the radiative decay of a meson A into
a meson B \cite{Close:2002ky,Bonnaz:2001aj},
\begin{eqnarray}
 A \to \gamma B\,.
\end{eqnarray}
In a non-relativistic quark model the transition amplitude, for a meson $A$ which
decays at rest to a meson $B$ plus a photon with momentum $p$, is given by the sum of
two contributions
\begin{eqnarray}
 {\cal M}_{A \to \gamma B} = {\cal M}^{(a)}_{A \to \gamma B} 
 + {\cal M}^{(b)}_{A \to \gamma B}\,,
\end{eqnarray}
The first contribution is related to the probability of emission of a photon by the
quark
\begin{eqnarray}
 {\cal M}^{(a)}_{A \to \gamma B} &=& \frac{\langle e_1 \rangle}{2 m_1} \int d^3p 
 \,\Phi_B^\ast \left( \vec{p} - \frac{m_2}{m_1 + m_2} \vec{k} \right) \nonumber \\
 && \times \left[2\vec{p} - i \sigma_1 \times \vec{k}\right] \Phi_A(\vec{p})
\end{eqnarray}
and the second one is related to the probability of emission of a photon by the
antiquark
\begin{eqnarray}
 {\cal M}^{(b)}_{A \to \gamma B} &=& \frac{\langle e_2 \rangle}{2 m_2} \int d^3p
 \,\Phi_B^\ast \left( \vec{p} + \frac{m_1}{m_1 + m_2} \vec{k} \right) \nonumber \\
 && \times \left[-2\vec{p} - i \sigma_2 \times \vec{k}\right] \Phi_A(\vec{p})
\end{eqnarray}
The two contributions for the amplitude are represented by the diagrams in 
Fig. \ref{diag1}.

The wave function for a spin 1 meson can be written in the following form
\begin{eqnarray}
 \Phi_M = \frac{1}{\sqrt{2}} {\cal Y}_{lm}(\vec{p}) \, \mathbf{\sigma}
 \, R_M(\vec{p})\,,
\end{eqnarray}
where $\mathbf{\sigma}$ is the Pauli matrix and $R_M(p)$ is the meson radial wave
function
\begin{eqnarray}
 R_M(q) = \exp{\left(-\frac{p}{2\beta^2_M}\right)}.
\end{eqnarray}
The decay width of $f_1(1285) \to \gamma V$ is given by
\begin{eqnarray}
 \Gamma(f_1 \to \gamma V) = \frac{8\alpha p E_V \beta^2 F^2 \langle e^2 
 \rangle}{3 m_{f_1} m_q^2} \left(1+ \lambda x +\frac{\lambda^2}{2} x^2\right)\,,
\end{eqnarray}
where $x=p^2/\beta^2$, the photon momentum is $p = m_{f_1}-m_V$, the vector
meson energy is $E_V \approx m_V$, 
\begin{eqnarray}
 \beta = \sqrt{\frac{2\beta_{f_1}^{2}  \beta_{V}^{2}}{\beta_{f_1}^2 + \beta_{V}^2}},
\end{eqnarray}
\begin{eqnarray}
 \lambda = \frac{\beta_{f_1}^2}{2(\beta_{f_1}^2 + \beta_{V}^2)}
\end{eqnarray}
and
\begin{eqnarray}
 F = \frac{\beta^4}{\beta_{f_1}^{5/2}  \beta_{V}^{3/2}} \exp{\left(
 -\frac{p^2}{8(\beta_{f_1}^2 + \beta_{V}^2)}\right)}.
\end{eqnarray}

The isospin factors for the radiative decay were shown by Close et al in Ref.
\cite{Close:2002ky}. Here we use the following values for these factors
\begin{eqnarray}
 \langle e^2 \rangle = \frac{1}{4}
\end{eqnarray}
for $n\bar{n} \to n\bar{n}$ with different isospin and
\begin{eqnarray}
 \langle e^2 \rangle = \frac{1}{9}
\end{eqnarray}
for $s\bar{s} \to s\bar{s}$. For more details about the calculation of this factors
see Ref. \cite{Bonnaz:2001aj}.

Since this is a non-relativistic model, we expect theoretical uncertainty because
the momenta of the particles in the final state are large. We estimate this uncertainty
by comparing the results of this model, obtained by Close et al \cite{Close:2002ky},
with the results of a relativistic model, obtained by Piotrowska et al 
\cite{Piotrowska:2017rgt}. We choose the results for the radiative decay of $\rho(1450)$
and $\omega(1420)$ to compare the models because these are the lightest states calculated
in both models (see Table \ref{tab3}).
\begin{table}[ht]
\caption{comparison between the results from Refs. \cite{Close:2002ky,Piotrowska:2017rgt}. 
\label{tab3}}
\begin{center}
\begin{tabular}{ccc}
\hline
Meson & Piotrowska et al. (keV) & Close et al. (keV) \\
\hline
$\rho(1450) \to \gamma \pi$ & $72 \pm 42$ & $61$ \\
$\rho(1450) \to \gamma \eta$ & $230 \pm 140$ & $106$ \\
$\rho(1450) \to \gamma \eta^\prime$ & $56 \pm 33$ & $61$ \\
$\omega(1420) \to \gamma \pi$ & $600 \pm 360$ & $510$ \\
$\omega(1420) \to \gamma \eta$ & $23 \pm 14$ & $11$ \\
$\omega(1420) \to \gamma \eta^\prime$ & $5 \pm 3$ & $5.7$ \\
\hline
\end{tabular}
\end{center}
\end{table}
We can see that the results obtained with the model used in this paper are in good
agreement with the results from a relativistic model. We expect the relativistic effects
on the radiative decay of $f_1(1285) \to \gamma\rho$ and $f_1(1285) \to \gamma\phi$ to
be smaller than in Table \ref{tab3} because the differences between the masses of the
initial and final states are smaller, which implies at a lower momentum for final state
particles.

Another possibility of uncertainty we have to discuss is the fact of the $\phi$ decays
to $\rho\pi$ $15\%$ of the time which could lead to a background for the $f_1(1285)
\to \gamma\rho$ decay rate. If we take this into account in our calculation, the
$f_1(1285) \to \gamma\rho$ decay will increase about $32$ keV, which implies less than
$2\%$ error. In addiction we expect in the experimental point of view the $\rho$ from the
$\phi \to \pi\rho$ decay can be distinguished of the $\rho$ from the $f_1(1285) \to
\gamma\rho$ by the presence of a pion associated to the vertex in the $\phi \to \pi\rho$
decay. All the processes which can contribute to the uncertainties has low branching
ratios which implies in small uncertainties.

We also investigate the uncertainties doe to the presence of a $n\bar{n}$ component in
the $\phi$ wave function. The mixing angle for the $n\bar{n}$ and $s\bar{s}$, given by
PDG is $\theta_V = 36.4^o$ \cite{Tanabashi:2018oca}. If we take this angle into account
the flavor coefficients are
\begin{eqnarray}
 \mid \phi \rangle = - c_1 \mid n \bar{n} \rangle - c_2 \mid s \bar{s} \rangle
\end{eqnarray}
where $c_1 = 0.014$ and $c_2 = 0.999$. If we consider the $n\bar{n}$ component of $\phi$
the decay rate must be multiplied by a factor of $c_2^2 = 2\times 10^{-4}$. At the same
time if we consider the $s\bar{s}$ component of $\phi$ the decay rate must be multiplied
by a factor of $c_1^2 = 0.9994$. Than we can see that if the $s\bar{s}$ component of 
$f_1(1285)$ is enlarged the $n\bar{n}$ component decreases and the glue content of
$f_1(1285)$ does not change. The same discussion is valid for the $\rho$ meson.
Even if the contribution of $n\bar{n}$ in $\phi$ was 10\%, this contribution would be
present in both our results and experimental data, implying that these contributions in
$f_1(1285)$ do not change. Thus, the contribution of gluons only changes in $f_1(1285)$
if $\rho$ and $\phi$ have contribution of gluons, which is not considered in the literature.

\section{Results}

The decay of $f_1(1285) \to \gamma \rho^0$ can be related to the content of quarks $u$
and $d$ doe to the $\rho^0$ is composed exclusively of these quarks. Our result for the
decay rate is
\begin{eqnarray}
 \Gamma\left(f_1(1285) \to \gamma \rho^0\right) = 1608\, \rm{keV}\,.
\end{eqnarray}
When we compare this result with PDG data for the same process ($1203 \pm 280$ keV)
\cite{Tanabashi:2018oca} we estimate the $u$ and $d$ content is about $75 \%$ of 
$f_1(1285)$. Which means that the $f_1(1285)$ is mostly composed by $u$ and $d$ quarks.
It is important to note that our results are also in agreement with the results of Close
et al \cite{Close:2002sf}. However, when we compare our result with the CLAS data
($453 \pm 177$ keV) the content of $u$ and $d$ on $f_1(1285)$ decreases to about of 
$28 \%$. Then we need to investigate the $s$ quarks content of $f_1(1285)$, on this way
we consider the decay $f_1(1285) \to \gamma \phi$. This decay channel give us
information about the $s$ component because the $\phi$ meson is constituted basically
of $s\bar{s}$ quarks. The result obtained is
\begin{eqnarray}
 \Gamma\left(f_1(1285) \to \gamma \phi\right) = 214\, \rm{keV}\,.
\end{eqnarray}
When we compare this result with PDG data for the same process ($17.0 \pm 6.3$ keV)
the content of $s$ quarks can be estimated as $8 \%$ of total constituents. The CLAS
collaboration do not have data about this channel. These results for $u$, $d$ and
$s$ quarks content of $f_1(1285)$ add up to $83 \%$ which imply in a glue content of
$17 \%$ in the $f_1(1285)$ when the PDG data was considered. If we consider the
uncertainty in the model we have used the results are inconclusive. But when we consider
the CLAS data the total contribution of $u$, $d$ and $s$ is about $36 \%$, which imply
in a large glue content of about $64 \%$. Even if the uncertainty in this model is about 
$50 \%$, we can still say that there is a significant content of gluons in $f_1(1285)$. 
These results confirm the results obtained by Birkel and Fritzsch \cite{Birkel:1995ct}
about the existence of a glue content for $f_1(1285)$. These results also imply the
existence of partner resonances with nearby masses in which a similar mixing will be
present like as $f_1(1420)$ and $f_1(1510)$ as proposed by Birkel and Fritzsch 
\cite{Birkel:1995ct}.  

\section{Summary and Conclusions}

In this paper we have calculated the radiative decay rates of $f_1(1285)$ to $\rho$ and
$\phi$ vector mesons. The choice of the $\rho$ and $\phi$ mesons is due to the fact that
they have a well-known internal structure and because there are experimental data for
these decay channel. The $\rho$ is exclusively composed by $u$ and $d$ quarks and the
$\phi$ meson is exclusively composed by $s$ quarks. The results were compered to
experimental data and this comparison allowed us to estimate the contribution of $u$,
$d$ and $s$ quarks for $f_1(1285)$ and consequently the glue content too. When compared
to the PDG data our results indicated that $f_1(1285)$ has a small or non existing glue
content. However, when our results were compared to the CLAS data the $u$, $d$ and $s$
quarks content decreased and consequently we found a large glue contribution to the
$f_1(1285)$ wave function.Even if the glue content of $f_1(1285)$ is small in a more
precise calculation it must be taken into account.

\section*{Acknowledgments}
The authors thank Victor P. Gon\c calves and Werner K. Sauter by comments and suggestions.
This study was financed in part by the Coordena\c c\~ao de Aperfei\c coamento de Pessoal
de N\'{\i}vel Superior - Brasil (CAPES) - Finance Code 001. This study was financed in part
by CNPq, Brazil.

\end{document}